\documentclass[twocolumn,prl]{revtex4}

\usepackage{epsfig}
\newcommand{\e}{{\rm e}}
\begin{document}

\bibliographystyle{apsrev}

\title{
Time-frequency bases for BB84 protocol
}

\author{S.~F.~Yelin}
\affiliation{Department of Physics, University of Connecticut, Storrs, CT 06269}
\author{Bing C.~Wang}
\affiliation{Department of Electrical and Computer Engineering, University of Connecticut, Storrs, CT 06269}

\date{\today}

\begin{abstract}
The bases traditionally used for quantum key distribution (QKD) are a 0 or $\pi/2$ polarization or alternatively a 0 or $\pi/2$ phase measured by interferometry. We introduce a new set of bases, i.e. pulses sent in either a frequency or time basis if the pulses are assumed to be transform limited. In addition it is discussed how this scheme can be easily generalized from a binary to an $N$-dimensional system, i.e., to ``quNdits.'' Optimal pulse distribution and the chances for eavesdropping are discussed. 
\end{abstract}


\maketitle


Quantum key distribution (QKD) was first proposed in its present form by Charles
H. Bennett and Gilles Brassard \cite{BB84}, and their protocol was
subsequently        called BB84.   QKD using the BB84 protocol uses
single photon pulses traveling between two users to create a
common encryption key. The             
sender, commonly called Alice, first randomly sends photons in two
non-orthogonal basis sets while randomly switches between the two
bases of each non-orthogonal set \cite{gisinrev}.   The receiver,
commonly called Bob, detects the incoming photons using a
random choice of basis. Subsequently, polarization encoding \cite{benn92c}, phase encoding \cite{benn92a}, pairs of entangled photons \cite{eker91,benn92b} based on EPR paradox \cite{EPR}, and phase shifting \cite{yama02} was proposed. In physical manifestations of
BB84 protocol, quantum bits,           also named
qubits, are most commonly represented by either the
polarization of light or the phase inside an optical interferometer.
Two non-orthogonal basis sets most commonly used are [0,$\frac{\pi}{2}$] and [$\frac{\pi}{4}$, $\frac{3\pi}{4}$], which can describes either the angle of polarization or the
relative phase between two arms of an interferometer.  Using
polarization or phase limits the number of orthogonal basis in each
set to two. Moreover, both traditional basis sets commonly used, i.e., a polarization basis and an interference basis (where a phase shift by $0$, $\frac{\pi}{2}$, $\pi$, or $\frac{3\pi}{2}$ would replace the polarizations) suffer from limitations in optical fiber. Polarization mode dispersion and temperature sensitivity of the optical fiber makes controlling the polarization and phase of single photon pulses over long distances difficult. Techniques used to mitigate the effect of polarization mode dispersion in optical fiber includes the plug-and-play technique cite{zbin97} and the auto-compensation technique \cite{beth00}.

This paper proposes a new QKD scheme, which introduces a
new set of bases and utilizes the uncertainty in information between
time and frequency domain.  Single photon pulses sent in the
time-basis set reside in different overlapping time bins. Similarly,
single photon pulses sent in the frequency-basis set reside in
different overlapping frequency bins.  Alice randomly transmits qubits
in either of the two basis sets.   One advantage to this approach is
that there is no limit to the number of basis used in either the
frequency or the time basis.  Though there are only two basis sets,
there can be an arbitrary number of time bins or frequency bins in
each set.   The ability to accommodate large number of bases in each
set allows the implementation of quNdit QKD system. 

In addition, since multiple non-overlapping frequency sets can
coexists in a larger frequency band, this scheme is also suitable for
applications in wavelength division multiplexed optical networks.

\begin{figure}[ht]
\epsfxsize=7cm
\centerline{\epsffile{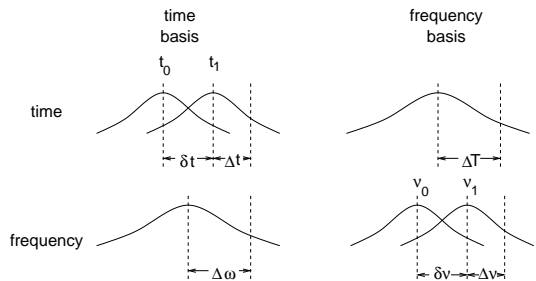}}
\caption{\protect\label{f_pulses}(a) Alice sends in ``time basis'': Two distinguishable pulses centered $t_0$, $t_1=t_0+\delta t$ and are $\Delta t$ broad. The same pulse has a broad frequency spectrum. (b) ``Frequency basis'': Here the roles are of time and frequency are changed, the broad time spectrum is centered around $(t_0+t_1)/2$ and has width $\Delta T$.}
\end{figure}

In the BB84 protocol ``0'' and ``1'' are $0^{\circ}$ and $90^\circ$ or
$45^\circ$ and $135^\circ$, depending on the basis. In the
time-frequency bases (Fig.~\ref{f_pulses}) ``0'' is a pulse sent at
time $t_0$, ``1'' one sent at $t_1$ if we are in the ``time basis.''
In this case, both have the same broad frequency spectrum centered at
the frequency $(\nu_0+\nu_1)/2$. If we are in the frequency basis
``0'' is a pulse centered at $\nu_0$, ``1'' one at $\nu_1$, while the
time-shape of both of these pulses would be the same, centered at
$(t_0+t_1)/2$.  

The following table  summarizes the binary values and the basis used by Alice to send single photons and what Bob measures using his choice of basis sets \cite{gisinrev}:
\begin{equation}
\begin{tabular}{c|c|c|c}
 Alice basis & Alice sends & Bob basis & Bob measures \\ \hline 
time & 0 & time & 0 \\
time & 0 & frequency & ? \\
time & 1 & time & 1 \\
time & 1 & frequency & ? \\
frequency & 0 & time & ? \\
frequency & 0 &frequency & 0 \\
frequency & 1 & time & ? \\
frequency & 1 & frequency & 1
\end{tabular}
\end{equation}
In order to understand this scheme, it is important to keep in mind that only {\it one} quantity of the single photon pulse can be measured: Quantum mechanics demands that the pulse can be measured only either in the frequency or in the time basis. (The same would be true for analogous schemes, e.g., one measures only the frequency, but uses pulse shape and phase as the two bases.)

The probability that Bob measures the wrong value even in the right basis is, of course, non-zero: Let's assume the time-width of the pulses in the time basis is $\Delta t$. (``Width'' means here the standard deviation.)
We also define $\delta t$ as the time separation between the two time bins used in the time basis, $\Delta T$ as the temporal width of the pulses sent in the frequency basis, and $\delta\omega$ as the spectral width of the pulses sent in the time basis. Then the following uncertainty equation governs the relationship between each time and frequency element: $\Delta T \cdot \Delta\nu = \Delta\omega\cdot\Delta t \ge 1$. 

Obviously, in order to not be able to distinguish the pulses in the ``wrong'' basis, 
\begin{equation}
\delta t\le \Delta T.
\end{equation}
The other relations can be understood from the following estimates:

Assume that Alice sends a pulse in the time basis. We assume normalized Gaussian pulses. (Note that all pulses must have the same area!)
If Bob's measurement consists of determining whether the pulse arrives before or after $(t_0+t_1)/2$, he measures correctly with the probability
\begin{equation}
\frac{1}{\sqrt{2\pi}\Delta t} \int\limits_{-\infty}^{t_0+\delta t/2} dt e^{-\frac{(t-t_0)^2}{2\Delta t^2}} \;=\; \frac{1+\Phi(x)}{2}\;,
\end{equation}
where $x=\delta t/2\sqrt{2}\Delta t$ and $\Phi$ is the error function. In order to have this probability larger than 99\%, $x$ would have to be at least 1.65. We call this quantity {\it fidelity}. 

\begin{figure}[ht]
\epsfxsize=9cm
\centerline{\epsffile{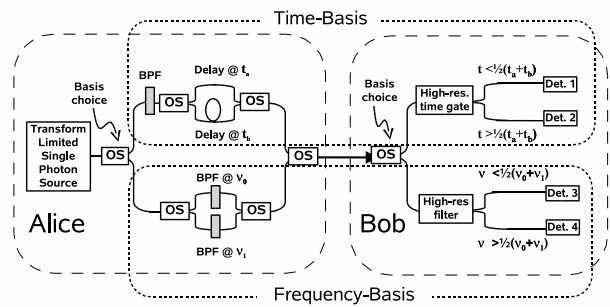}}
\caption{\protect\label{f_exp}The setup of the proposed time-frequency
  QKD bases. The lines in the figure represent optical fibers. OS:
  Optical Switch. BPF: Optical 
Band Pass Filter; Det.: Single photon detectors.}
\end{figure}

Figure \ref{f_exp} shows an experimental setup that can implement the
time--frequency QKD bases using commercially available fiber optic
components.   The setup in Fig.~\ref{f_exp} is first divided into
Alice, contained in the left box, and Bob, contained in the right box.
The setup is also divided into the time basis, contained in the top
box, and frequency basis, contained in the bottom box.  In the box on
the left, Alice injects photons from a transform-limited single photon
pulse source into an electro-optic switch.   The optical switch
randomly switches the photon between the time basis and the frequency
basis.  In the time basis, the pulse first passes through a band pass
filter (BPF), and then an optical switch that randomly switches the
photon between two delay values, around $t_0$ and $t_1$.  In the
frequency basis, the pulse passes through an optical switch that
randomly switches the photon between two BPF centered at $\nu_0$ and
$\nu_1$.   The output from Alice's transmitter in both the time and
the frequency basis are combined and transmitted to Bob.   

Once the single photon pulses arrive at Bob, he randomly detects the
photon using either the time basis or the frequency basis.  In the
time basis, the photon enters a high resolution time gate that will
place photons arriving at times $t < \frac{1}{2} (t_0 + t_1)$ and $t >
\frac{1}{2} (t_0 + t_1)$ into separate optical fiber output ports.  In
the Frequency Basis, the photons enter a high resolution optical
spectral filter that will place photons with frequencies $\nu <
\frac{1}{2}(\nu_0 + \nu_1)$  and $\nu > \frac{1}{2}(\nu_0+\nu_1)$ into
separate optical fiber output ports. Like the original BB84 protocol,
Bob records all photons detected at each of the four photon detectors
but discards the photons where Bob and Alice use incompatible bases.
The photons detected when the same basis are used by Alice and Bob
will be the ones actually used for encryption key distribution.  These
photons are also known as sifted key.  

Security against eavesdropping is strongly based on the so-called
no-cloning theorem \cite{noclone} that makes it impossible to copy a
qubit. 
We therefore assume an intercept-resend strategy for the eavesdropper
(Eve). That is, her best strategy is to intercept photons between
Alice and Bob, measure them, and resend a photon pulse according to
what she measured. 

If the Eve measures in the same way as Bob does there is no difference
to the traditional BB84 protocol. She can, however, do something that
might be smarter in this case. She could just measure only a small
slice around $t_0$ and $t_1$, let's say, in the slices
$t_0-\Delta\tau$ to $t_0+\Delta\tau$, and the same around $t_1$. If
the pulses in the time basis are considerably narrower than the time
width of the pulse if sent in the frequency basis (i.e. $\Delta T$),
and she measures something in, let's say, the $t_0$ slice, then she
can be sure enough that it was indeed sent in the time basis as a 0
that she can store this information, ``0,'' and send herself a time
basis pulse centered around $t_0$ on to Bob. In this case the
probability that she is wrong is small. Here we quantify this train of
thought: 

If the pulse was sent in the time basis, the fraction of the pulse and
therefore the probability for Eve to measure the right photon, is 
\begin{equation}
\label{eve1}
p_1 \;=\; \frac{1}{\sqrt{2\pi}\Delta t}
\int\limits_{t_0-\Delta\tau}^{t_0+\Delta\tau} dt \,
e^{-\frac{(t-t_0)^2}{2\Delta t^2}} = \Phi(y) \approx
\frac{2}{\sqrt{\pi}} y\;, 
\end{equation}
where $y=\Delta\tau/\sqrt{2}\Delta t$. The probability that she
measures th wrong photon is 
\begin{eqnarray}
p_2 &=& \frac{1}{\sqrt{2\pi}\Delta t}
\int\limits_{t_0-\Delta\tau}^{t_0+\Delta\tau} dt \,
\e^{-\frac{(t-t_1)^2}{2\Delta t^2}} \;=\\ 
&=& \frac{1}{2} \left( \Phi(2x+y)-\Phi(2x-y) \right) \;\approx\;
\frac{2\e^{-4x^2}}{\sqrt{\pi}}y\;.\nonumber 
\end{eqnarray}
If, on the other hand, Alice sent a pulse in the frequency basis the
probability Eve measures a photon at all is proportional to the
overlap of one of her two slices with the broad time profile: 
\begin{eqnarray}
\label{eve2}
p_3 &=& \frac{2}{\sqrt{2\pi}\Delta T}
\int\limits_{t_0-\Delta\tau}^{t_0+\Delta\tau} dt \;
\e^{-\frac{(t-t_0-\delta t/2)^2}{2\Delta T^2}} \;=\\ 
&=&  \Phi\left(\frac{x+y}{z}\right) - \Phi\left(\frac{x-y}{z}\right)
\;\approx\; \frac{4}{\sqrt{\pi}} \e^{-\frac{x^2}{z^2}} \frac{y}{z}
\;,\nonumber 
\end{eqnarray}
where $z=\Delta T/\Delta t$. 

Assume Eve is sending each ``result'' she gets to Bob. Then there are
two questions: (i) Which percentage of the key would she get? (ii)
What additional percentage of error would she introduce into the
transmission from Alice to Bob? 

The answer to question (i) can be found by considering the following:
We only have to consider the cases where Alice and Bob use the same
basis. Then, in half of the remaining cases, Eve uses the same basis,
too, whereas in the other half she guesses 50\% right. The total part
of the sifted key that Eve gets correctly is thus given by 
\begin{eqnarray}
P &=& \frac{1}{2} \Phi(y) + \frac{1}{2} \cdot\frac{1}{2} \left(
  \Phi\left(\frac{x+y}{z}\right) -
  \Phi\left(\frac{x-y}{z}\right)\right) \;\approx\nonumber\\
&&\frac{y}{\sqrt{\pi}} \left(1 + \frac{\e^{-\frac{x^2}{z^2}}}{z} \right)
\;. 
\end{eqnarray}

The second important quantity to be found is the additional error she
introduces, since that serves as a measure for Alice and Bob to detect
an eavesdropper. Again, we consider the two cases: In the half of the
cases where the bases coincide, Eve introduces an error when she
measures the photon wrong, therefore resends it wrong, but the Bob
measures ``correctly'' the wrong photon. In the other case, Eve
resends in the wrong basis, where Bob ``guesses'' the wrong value with
probability $1/2$. The total additional error is thus 
\begin{eqnarray}
E &=& \frac{1}{2} \cdot\frac{1}{2}
\left(\Phi(2x+y)-\Phi(2x-y)\right)\frac{1+\Phi(x)}{2} + \nonumber \\
&& \frac{1}{2}
\cdot \frac{1}{2} \left( \Phi\left(\frac{x+y}{z}\right) -
  \Phi\left(\frac{x-y}{z}\right) \right) \;\approx\\ 
&=& \frac{y}{\sqrt{\pi}} \left( \e^{-4x^2} \frac{1+\Phi(x)}{2} +
  \frac{\e^{-\frac{x^2}{z^2}}}{z} \right)\;. \nonumber 
\end{eqnarray}

The ratio of the two, $R=E/P$, can serve as a figure of merit, since
Alice and Bob try to minimize $P$ while at the same time trying to
maximize $E$. 

For a fidelity of 99\%, $x\approx 1.65$ and $z$ in the range between 2
and 3 this ratio would be $R\approx5$, i.e., Eve would get $0.7y$ of
the key correctly, while she introduces an additional error of
$0.15y$. This ratio of $R=5$ has to be compared to a ratio of 2 in the
case of the traditional polarization scheme. The ratio in our case,
depends strongly on the desired fidelity. If we are content, e.g.,
with a fidelity of 90\%, the ration $R$ drops to a value slightly
above 2. Obviously, the broader the slices are that Eve measures, the
closer she comes towards the ratio of $R=2$, no matter what the
fidelity is.

\begin{figure}[ht]
\epsfxsize=7cm
\centerline{\epsffile{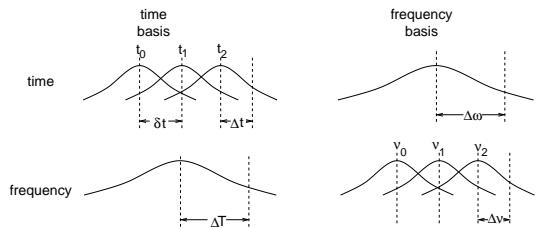}}
\caption{\protect\label{f_three} Same as in Fig.~\ref{f_pulses}, but
  for the case of qutrits, where a pulse centered around $t_0$ means
  ``0'', etc.} 
\end{figure}

The time-frequency basis also lends itself easily to generalization to
quNdits, i.e., states in $N$-dimensional Hilbert space. In this case,
$N$ possible pulses would be centered around $t_0, t_1, \ldots
t_{N-1}$ for $N$ different bases in the  time basis or $\nu_0, \nu_1,
\ldots \nu_{N-1}$ for $N$ different frequency bins in the frequency
basis. Intuitively, the total difference between the first and the
last pulse should be of the same order as the total pulse width in the
wrong basis (see prove of this statement in \cite{crem03}). For
qutrits, the pulse scheme is depicted in Fig.~\ref{f_three}. 

One drawback of this extension in that increasing the number of bases will also improve the result of Eve's attack strategy. The ratio of the likelihoods to measure a photon around $t_i$ in the right and wrong basis goes down, since the total pulse areas have to remain constant. Thus the ratio $R$ increases linearly with increasing $N$ \cite{crem03}.

In conclusion, we propose a new quantum key distribution scheme that is based on  the quantum mechanical duality between time and frequency. We described an experimental implementation of the new scheme using optical fiber and discuss the scheme's security against eavesdropping. The extension to an $N$ basis representation of quNdits provides the new scheme with an advantage over traditional implementations of the BB84 protocol. However, this advantage is somewhat mitigated by the fact that increasing the number of bases also improves Eve's ability to eavesdrop on the key distribution process.

We would like to acknowledge helpful discussions with R. Cremona and M. Kostrun. The project is supported by National Science
Foundation Information Technology Research Grant Number EIA-0312890.
 
\bibliography{timefreq_s}

\end{document}